# In Vitro Evaluation of Cytotoxic and Anti-HCV-4 Properties of Sofosbuvir Encapsulated Chitosan Nanoparticles

Samah A. Loutfy[1,*], Hosam G Abdelhady [2,3] , Mostafa H. Elberry[1], Ahmed R. Hamed[4], Hussien Ahmed[5] , M T M Hasanin[6], Ahmed Hassan Ibrahim Faraag[7], El-Chaimaa B. Mohamed[1], Ashraf E. Dardeer[1], Reham Dawood[8], Yasmin Abo-zeid[9], Mostafa El-Awady[8]

[1] *Virology and Immunology Unit, Cancer Biology Department, National Cancer Institute, Cairo University, Egypt.*

[2] *Department of Pharmaceutics and pharmaceutical technology, College of Pharmacy, Taibah University, Al Madinah Al Munawwarah, Saudi Arabia.*

[3] *National Organization for drug control and Research, Cairo, Egypt*

[4] *Phytochemistry Department & Biology Unit, Central Laboratory of the Pharmaceutical and Drug Industries Research Division, National Research Centre, El-Bohuoth st. (formerly El-Tahrir), Dokki 12622, Giza, Egypt.*

[5] *Faculty of Medicine, Zagazig University, Zagazig, El-Sharkia, Egypt.*

[6] *Nanotech Dreamland, El-Wahaat Road, 6th October, Giza, Egypt.*

[7] *Botany and Microbiology Department, Bioinformatics Center, Faculty of Science, Helwan University*

[8] *Department of Microbial Biotechnology, National Research Centre, Giza, Egypt.*

[9] *Pharmaceutics Department, School of Pharmacy, Helwan University, Cairo, Egypt.*

* Corresponding author: Samah Aly Loutfy, Ph.D., Virology and Immunology Unit, Cancer Biology Department, National Cancer Institute, Fom El-Khalig, Cairo 11796, Egypt. Tel.:+20122840964. Email: samaly183@yahoo.com

## Abstract

Sofosbuvir is a potent HCV NS5B nucleotide polymerase inhibitor with broad genotypic coverage and low risk of developing drug resistance. While clinical studies have provided the effectiveness of sofosbuvir for treatment of patients with hepatitis C virus genotype 4 (HCV-4), however, many side-effects were reported. To reduce those side effects and improve the antiviral activity of the drug, sofosbuvir was encapsulated into chitosan nanoparticles (CNPs) to produce sofosbuvir encapsulated chitosan nanoparticles (SCNPs). 3D Molecular simulation and dynamics were made for sofosbuvir and SCNPs to evaluate the active sites of HCV-4 NS3 protease, NS5B polymerase and HCV helicase relative to both their catalytic activities and drug susceptibilities. The produced SCNPs were finally evaluated on hepatoblastoma cells (Huh7) for their antiviral efficiency.

**Methods:** SCNPs were prepared by ionic gelation method and characterized by FTIR, TEM, Zeta sizer and U-Vis Spectrophotometry. The cytotoxicity of SCNPs were then performed on Huh7 cells using flow cytometry and colorimetric MTT techniques. The anti HCV-4 activity of SCNPs was determined by quantitative real time PCR. Finally, the 3 D structure molecular modeling and simulation were made by Molegro Virtual Docker software and Swiss model server respectively.

**Results:** The suggested chemical structure for SCNPs was confirmed by FTIR. The average particle size and surface charge of SCNPs were 137 ± 34 nm and 29 ± 9.6 mV respectively. The Encapsulation efficiency was 80% and the loading efficiency was 6%. The binding affinity of SCNPs with HCV-4 NS3, NS5B and NS5A were -156.512, -154.603 and -131 respectively. But for sofosbuvir were -127.581, -131.535 and -167 respectively based on the MolDock score. The treatment of HCV- 4 infected Huh7 cells with up to 100 µM of SCNPs neither showed significant cytotoxic nor genotoxic effects after 48 hours of cell exposure. Finally, a complete disappearance of HCV RNA was seen after 24 hours of exposure to 100 μM of SCNPs when compared with the untreated cells.

**Conclusion:** The SCNPs displayed potent anti HCV-4 activity in Huh7 cells with no apparent cellular toxicity. Molecular docking could be a promising tool for predicting novel antiviral agents in saving time, effort and cost before applying in vitro assays. We believe that this study provides promising application of CNPs as a safe nanocarrier for antiviral agents.

**Keywords:** Sofosbuvir, Chitosan nanoparticles, Direct acting antiviral agents, cytotoxicity

# 1   Introduction

Viral infections represent a major public health issue, with negative impacts not only on healthcare but also have numerous socioeconomic costs. Although, there are many antiviral agents available in the market with different antiviral mechanisms, viruses developed resistant to most of them. This is mainly due to their selective action against one virus or viruses belong to the same family as well as the absence of broad-spectrum antiviral agents. Therefore, there is a high demand for discovery of novel strategies to improve the antiviral therapies to control or limit the spread of infections.

Hepatitis C is considered as a major problem worldwide with more than 200 million people infected globally. Egypt has the highest prevalence of hepatitis C (1). Hepatitis C infection is called the silent killer as it could infect the patients for long time without appearance of any symptoms until late stages of disease. Patient with chronic hepatitis suffer from weakness and jaundice with a possibility to extend to liver cirrhosis and hepatocellular carcinoma leading to death (1). Efficient treatment of hepatitis C infections with conventional antiviral agents is hindered by the development of resistance to antiviral agents due to virus mutation (2), and adverse side effects due to drug accumulation at off-target organs (3).

Most recently, direct acting antiviral agents (DAA) such as protease, polymerase, NS5B or NS5A inhibitors were co-administered with ribavirin (double therapy) or with ribavirin and interferon (triple therapy) for treatment of hepatitis C virus (HCV) (4,5). Sofosbuvir (Sovaldi) is one of the most recent DAA that inhibits the replication of HCV through the inhibition of NS5B. However, some major side effects, especially in kidney and liver, were reported when sofosbuvir were used clinically in Egyptian patients.

This creates a huge pressure to develop novel antiviral agents to overcome these drawbacks. Nanoparticles (NPs) are considered as one of the most effective and novel therapeutic agents due to their potential ability to be tolerated and to target the therapeutic agent to the diseased organ with a concomitant decrease of off-target side effects (6).

Chitosan is a naturally occurring cationic polysaccharide possesses mucoadhesive properties that enable its transport across the mucosal membrane and known to be biodegradable and compatible with biological system due to its biodegradation into nontoxic amino sugars that can be completely absorbed by the body (7). Chitosan nanoparticles has been reported to encapsulate several drug efficiently and target their delivery into the affected organ (7,8).

In this work, we hypoethsis that sofosbuvir as a model of DAA when encapsulated into chitosan nanoparticles, might be associated with an increase of its safety profile and improve its antiviral activity in vitro in Huh7 cells infected with HCV – 4 due to the potential delivery of the drug into the affected organ (passive targting) e.g. liver. Consequently, there is a high possibility of reducing dose and the frequency of administration required for a proper therapeutic management of infected patients.

## 2. Materials and Methods

**2.1. Materials:** Chitosan (low molecular weight), tripolyphosphate (TPP), ethanol, methanol, ethyl acetate, glutaraldehyde, and isopropanol were purchased from Sigma-Aldrich (St Louis, MO,

USA). MTT [3-(4,5- dimethylthiazol-2-yl)-2,5-diphenyltetrazolium bromide] reagent was purchased from Serva (Heidelberg, Germany). Dulbecco's modified Eagle's medium (DMEM), penicillin, streptomycin, and fetal bovine serum (FBS) were purchased from Gibco (Merelbeke, Belgium). Chitosan (low molecular weight), tripolyphosphate (TPP), ethanol, methanol, ethyl acetate, glutaraldehyde, and isopropanol were purchased from Sigma-Aldrich (St Louis, MO, USA). MTT [3-(4,5- dimethylthiazol-2-yl)-2,5-diphenyltetrazolium bromide] reagent was purchased from Serva (Heidelberg, Germany). Dulbecco's modified Eagle's medium (DMEM), penicillin, streptomycin, and fetal bovine serum (FBS) were purchased from Gibco (Merelbeke, Belgium). The Huh7 cells were supplied by Rockefeller University (New York, NY, USA). Sofosbuvir was a kind gift from Gilead company.

## 2.2. Methodology

### 2.2.1. Preparation of Sovaldi -chitosan nanoparticles (SCNPs)

SCNPs was prepared by following the previously reported protocol (9). Briefly, sofosbuvir powder was dissolved in 3.8 ml of dimethyl sulfoxide to form 5 mM solution. Sofosbuvir solution was then added to the chitosan solution (1% acetic anhydride solution, 5 mg/ml, 100 ml) and mixed by continuous stirring for 1 h. The mixture was then neutralized by the addition of the negatively charged tripolyphosphate (TPP) solution (distilled water, 2.5 mg/ml, 67 ml), under magnetic stirring (500 rpm, 1 h) and at room temperature (11). The obtained NPs were then washed with distilled water (50 ml X 5) followed by centrifugation at 15,000 rpm to remove the residual acetic acid and other impurities. The produced pellets were re-suspended in distilled water by the help of vortex.

## 2.2.2. Characterization of SCNPs

### 2.2.2.1. Transmission electron microscopy (TEM)

SCNPs were imaged using TEM (H-700, Hitachi Ltd., Japan) at accelerated voltage of 80 KV. TEM was performed for identification of the sample uniformity, presence of aggregation (if any), shape, and size. After samples were applied on TEM grid, the grid was left to dry in air before imaging.

### 2.2.2.2. Particle size and zeta potential:

The particle size and their zeta potential were measured by zeta sizer (Malvern Instruments ZS NANO, UK). Samples were diluted in distilled water to give count rates ranging from 50 to 300 Kcps and the measurements were performed at 25 °C ± 0.1(7).

### 2.2.2.3. Encapsulation efficiency and drug loading

The encapsulation efficiency was determined by measuring the sample absorbance at $ʎ_{max}$ 260 nm using U-Vis Spectrophotometry (V-650 UV-VIS Spectrophotometer, Jasco, Japan). A calibration curve covering a concentration range of 0.5 to 5mM was constructed to calculate the encapsulation efficiency and the drug loading percentage using the following equations;

$$\text{Encapsulation Efficiency\%} = \frac{Weight\ of\ drug\ encapsulated}{weight\ of\ initial\ drug} * 100$$

$$\text{Drug Loading \%} = \frac{weight\ of\ Drug\ encapsulated}{weight\ of\ polymer\ used\ to\ form\ nanoparticles} * 100$$

### 2.2.2.4. Fourier-transform infrared spectroscopy (FTIR)

The FTIR was carried out for CNPs, Sovaldi, and SCNPs to show the interaction bonds between sovaldi and chitosan. FTIR was used with the following specifications: spectral range 400–4500

cm−1, number of scans 16, spectral resolution 4 cm−1. The software was Opus (Bruker (Germany), VERTEX 70, RAMII).

### 2.2.3. Cytotoxicity study of SCNPs in vitro

#### 2.2.3.1. Cell Culture

The human hepatoblastoma cell line, Huh 7 cells (ATCC, USA) was cultured by following the previous reported protcol (10) using Dulbecco's Modified Eagle Medium (DMEM, Biowest, France) containing fetal bovine serum (10% Biowest, France) and antibiotics (100 IU/ml penicillin/streptomycin, Lonza, Belgium) as a growth medium. Cells were then maintained in a monolayer culture at 37ºC under a humidified atmosphere and 5% $CO_2$. Cells were routinely subcultured by trypsinization (Trypsin/versine, Lonza, Belgium).

#### 2.2.3.2. Microscopic study

Huh7 Cells ($3x10^6$, 75cm$^2$flask) (Corning, USA) were treated with SCNPs (100 µM) for 24 h. Cells were then washed with PBS buffer, fixed with 2% glutaraldehyde for 2 h, then washed twice with PBS and finally fixed in 1% $OsO_4$ for 1 h. Following agarose (1.5%) enrobing, Spurr's resin embedding, and ultrathin (50 nm) sectioning, the samples were stained with aqueous uranyl acetate (2% w/v) and lead citrate (25 mg/ml) and imaged with a JEOL 100S microscope (12). Another set of cells was prepared similary but with no treatment with SCNPs to act as Control.

#### 2.2.3.3. MTT assay to identify cell viability of cells treated with SCNPs

We employed a modified method utilizing MTT (3-[4,5-dimethylthiazol-2-yl]-2,5-diphenyltetrazolium bromide dye (Serva Electrophoresis GmbH, Germany) that is based on the reduction of the dye by mitochondrial dehydrogenases of metabolically active cells into insoluble formazan crystals (10,13). Briefly, cell monolayers were treated in octuplet with control or test samples (sofosbuvir and SCNPs) for an exposure time of 48 h. At the end of exposure, MTT solution in phosphate buffered saline (PBS) (5 mg/ml) was added to each well and incubated for 90 min. The formation of formazan crystals was visually confirmed using phase contrast microscopy. DMSO (100 µl/well) was added to dissolve the formazan crystals with shaking for 10 min. The absorbance was measured at 540 nm against Blank (media that doesn't contain any cells) on a FLuo Star Optima microplate reader (BMG Technologies, Germany). Cell viability was calculated by comparing the OD values of the DMSO control wells with those of the samples and expressed as % viability to the control. Dose-response experiment was performed on samples producing $\geq 50\%$ loss of cell viability using five serial 2-fold dilutions (100, 50, 25, 12.5 and 6.25 µM) of the sample. Where applicable, $IC_{50}$ values (concentration of sample causing 50% loss of cell viability of the control i.e. untreated cells) were calculated using the dose response curve fit to non-linear regression correlation using GraphPad Prism® V6.0 software.

$$\% \text{ Viability of Huh7 cells} = \frac{\text{(mean OD}_{540}\text{ of test sample)}}{\text{(mean OD}_{540}\text{ of control)}} \times 100$$

### 2.2.3.4. TEM and cellular uptake

A similar set for cell treated with SCNPs (100µM) in section 2.2.3.3 was prepared but with no treatment with MTT reagent and these cells were imaged by TEM to identify the interaction and cellular localization of SCNPs in Huh 7 cells.

### 2.2.3.5. Flow cytometry study to measure the effect of SCNPs on cell cycle of Huh7 cells

Huh7 cells ($5 \times 10^5$ cells/well) were plated in 6-well microplates. After treatment with SCNPs (100 µM), cells were trypsinized, washed twice with PBS, suspended in PBS (300 µl) and finally fixed with ice-cold 70% ethanol (4 ml). To stain with propidium iodide (PI), cell sedimentation was performed by centrifugation; the ethanol was removed and washed with PBS. The cell pellets were then resuspended in 1ml of PI/Triton X-100 staining solution (0.1% Triton X-100 in PBS, 0.2 mg/ml RNase A, and 10 mg/ml PI) and incubated for 30 min at room temperature. The stained cells were analyzed using a MoFlo flow cytometer (DakoCytomation, Glostrup, Denmark) to calculate the fractions of cells in each phase of the cell cycle (14) versus another set of cells that was prepared similarly but were not treated with SCNPs to act as a control.

### 2.2.3.6.Identify DNA fragmentation for cells treated with SCNPs

Fragmentation of cellular DNA was investigated following the treatment of Huh7 cells with Sovaldi, SCNPs (100 μM) for 24 h. A fixed amount (100 ng) of cellular DNA (Genomic DNA Purification Kit, Amersham Biosciences) extracted from SCNPs treated and untreated cells, were resolved on 1.5% agarose gel electrophoresis in Tris-acetate buffer pH 8.2 and stained with 0.5 μg/ml ethidium bromide (EB). The DNA bands were examined under UV transillumination and photographed. Smearing, or presence of many low molecular weight DNA fragments are characteristic features of apoptotic cells (15).

### 2.2.4.   Antiviral activity of SCNPs

### 2.2.4.1. Viral inoculation in Huh7

Huh 7 Cells ($10^6$) were grown in 25 cm$^2$ flask for 48 hr to semi-confluence in DMEM containing 10% FBS, then, medium was removed and cells were washed twice with FBS-free medium, then inoculated with a serum sample  (500 μL sense) obtained from HCV-4 infected patients (RT-PCR and antibody positives) followed by addition of 500 μL of FBS-free DMEM. The HCV genotype in the applied sera was previously characterized as HCV-4 based on the method described by Ohno and his colleagues (16). The viral load was quantitated by real-time PCR and the average copy number was found to be $10^6$ IU/ml. After 90 min, DMEM containing FBS was added to make the overall serum contents around 10% in a final volume of 8 ml including the volume of human serum used for infection as mentioned above. Cells were maintained overnight at 37ºC and 5% $CO_2$. Adherent cells were then washed, three times, with culture medium to get rid of the remaining infection serum followed by addition of DMEM containing 10% FBS and cells were incubated for 48 h. Cells were harvested and assessed for the presence of HCV core proteins (17). Finally, quantitative PCR was performed (18).

### 2.2.4.2. Antiviral assay of SCNPs in Huh7

This was performed according to the previously published protocol by Hu and teamwork (17) and El Awady and his colleagues (19). Six well plates were seeded with 5x $10^5$ cells/well. After 24 h, A 500 µl of the positive serum was then inoculated and kept for 90 min for virus adsorption onto cells. DMEM containing 10% FBS (2.5 ml) was added followed by incubation at 37 ˙C and 5% $CO_2$ for 24 h. Infected cells were treated with either sovaldi or SCNPs. Untreated infected cells ( Control) represent cells that are cultured similarly but with no Sovaldi treatment. The plates were then subjected to RNA extraction and real-time PCR assay according to the previously published protocol (20).

### 2.2.5. In silico docking study of HCV-4 NS3, NS5A and NS5B Polymerase with Sovaldi and SCNPs

The docking study was carried out on an Intel Dual-Core (1.8 GHz) Windows PC with Molegro Virtual Docker (MVD) 2013.6.0.0 (15). The two-dimensional structure of Nano-chitosan was converted to three-dimensional format for docking purposes using Schrödinger's Materials Science Suite V2.6. Sovaldi ligand was obtained from PubChem Bioassay. In the absence of a crystal structure, homology models of NS3 protease and NS5A polymerase were built based on the published cocrystal structure of NS3 protease and NS5A polymerase using SWISS-MODEL protein structure homology-modelling server (16–19). The crystallographic structure of NS5B polymerase was obtained from the protein data bank (4wtg pdb).

## 3.     Results

### 3.1. Preparation and characterization of SCNPs:

SCNPs were prepared by ionic gelation method following a previous reported protocol as described in section 2.2.1. **Figures 1** showed TEM images of SCNPs, nanoparticles are spherical in shape and they are stable as revealed by the absence of aggregation. The particle size was presented as diameter ± Standard deviation and it was 137 ± 34 nm while zeta potential of SCNPs was 29 ± 9.6 mV. The encapsulation and loading percentage of Sovaldi were found to be 80% and 6% respectively.

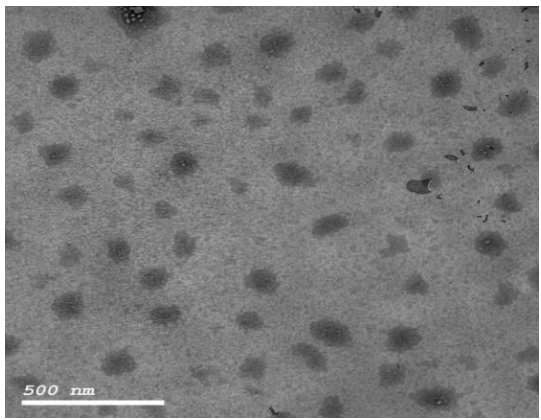

**Figure 1:** TEM images of sovaldi-chitosan nanoparticles

**FTIR analysis** showed the FT-IR spectra of the prepared CNPs and SCNPs. FTIR analysis depicts important absorption bands to identify the prepared CNPs. The stretching vibration of –OH bond was seen at 3430 cm$^{-1}$ while the absorption peak at 2929cm$^{-1}$ was appointed to a C-H stretching.

Finally, the absorption peaks at 1633 cm$^{-1}$, 1543 cm$^{-1}$ and 1385 cm$^{-1}$ were associated with the presence of C=O stretching of the amide band, bending vibration of N-H and bending vibration of O-H bond respectively. FTIR analysis showed the disappearance of the N-H peak and this indicative of the interaction between chitosan and sofosbuvir through the positively charged ammonium group of chitosan and the lone pair of electrons of sofosbuvir.

### 3.1. Cytotoxicity test

### 3.1.1. Light microscopy

**Figure 2 (A, and B)** depict light microscopy images of the control and SCNPs (100 µM, 53 mg/ml) treated Huh7 cells respectively. The treated cells were exposed to SCNPs for 24 hr. Interestingly, SCNPs showed no profound morphological changes or apparent cytotoxic effects.

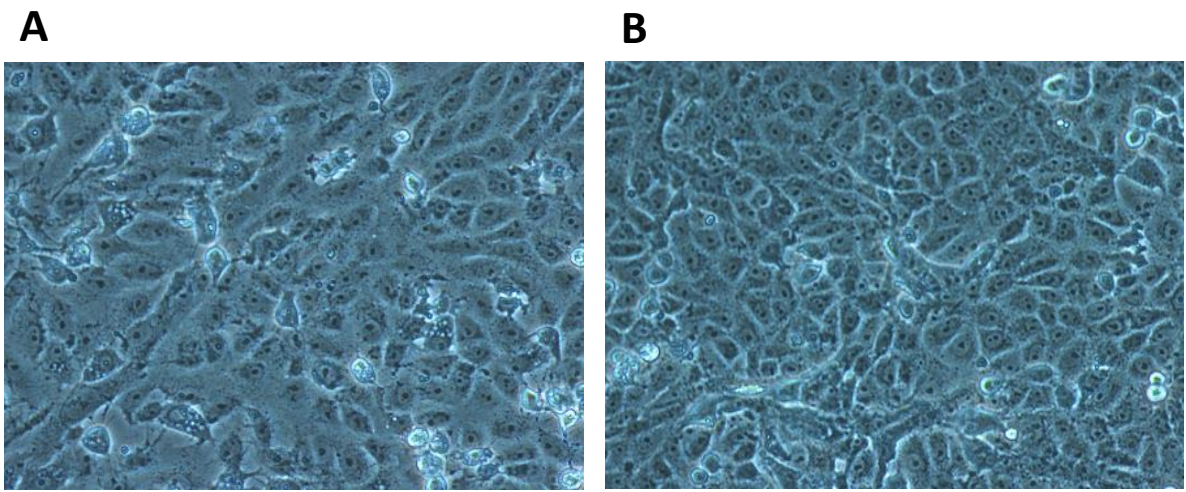

**Figure 2:** microscopic examination of Huh7 cells (A) Untreated and (B) treated with sovaldi-chitosan nanoparticles (100µM)

### 3.1.2. Dose-dependent cytotoxicity effect of SCNPs on HepG2 cells

**Figure 3 A and B** depict the cytotoxic effect of various concentrations of Sovaldi and SCNPs (5µM, 12.5 µM, 25µM, 50 µM and 100 µM) on the viability of Huh7 cells as determined by MTT

colorimetric assay after 48 h of exposure. Both Sovaldi and SCNPs, at concentrations up to 100 µM (53 mg/ml), did not reveal any cytotoxic effect over 48h. However, as revealed from **Figure 3A and B**, cells treated with SCNPs proliferate more than those treated with Sovaldi as indicated by the percentage of cell viability.

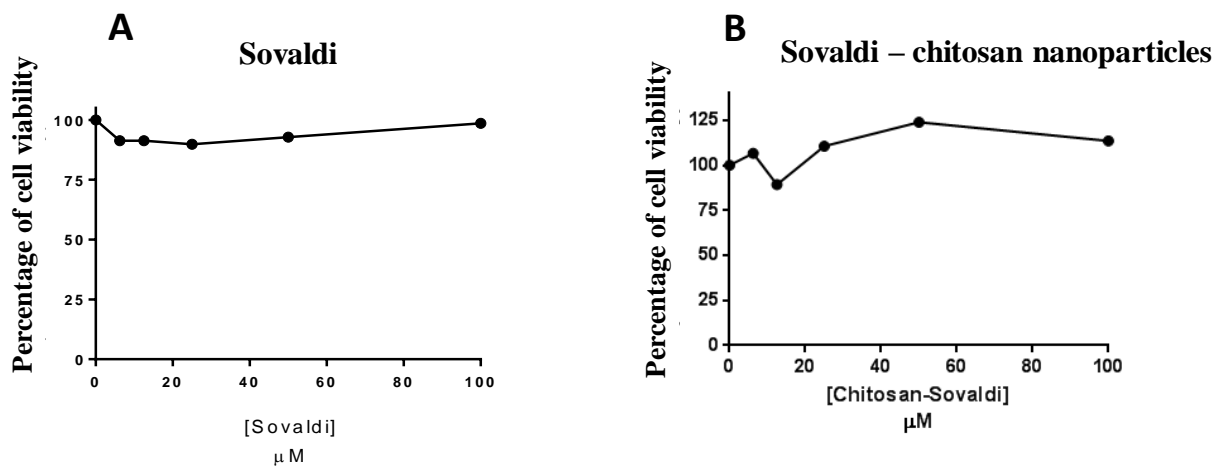

**Figure 3:** The effect of different concentrations of (A) Sovaldi and (B) Sovaldi – chitosan nanoparticles on the viability of Huh7 cells

### 3.1.3. Localization of SCNPs in Huh7 Cells by TEM

**Figure 4** shows the localization of SCNPs in Huh7 as seen by TEM. Binding and internalization of SCNPs in Huh7 cells was identifies where clusters of SCNPs were seen attached onto the cell membrane confirming cellular uptake of these NPs. Examination of images at higher magnification power showed intracellular NPs clusters, mainly associated with the internal membranes, with most of the dispersed NPs in the cytoplasm.

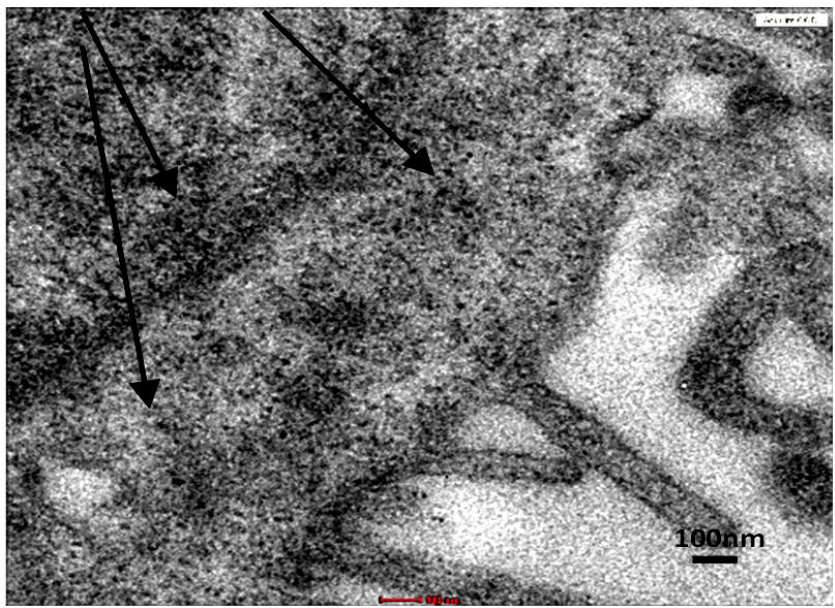

**Figure 4: TEM images showed the localization of sovaldi- Chitosan nanoparticles in Huh7 Cells**

### 3.2. Effect of SCNPs on cell cycle analysis

Cells treated with 100 µM (53 mg/ml) of SCNPs were evaluated for cell cycle and DNA contents using flow cytometric analysis. Data presented in **Table 1 and Figure 5** revealed similar cell cycle patterns of cellular growth in both treated and untreated cells with no obvious harmful effect was seen on the cellular viability after treatment with SCNPs.

**Table 1: Cell cycle analysis of Huh7 cells treated with SCNPs versus untreated cells (control)**

| Cell Cycle | G0-G1 | G2-M | G2/G1 | S | Diploid |
|---|---|---|---|---|---|
| **Control** | 71.07 % | 4.98 % | 2 | 23.95 % | 100% |
| **SCNPs (100 µM)** | 68.72% | 5.5% | 2 | 25.78% | 100 % |

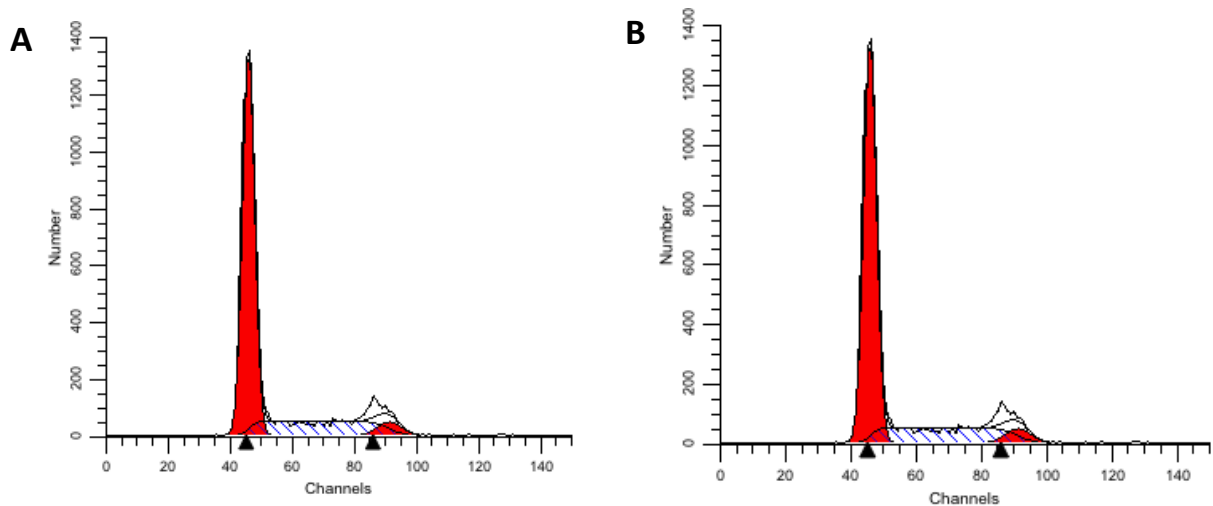

**Figure 5: Flow cytometry study showed cell cycle analysis of Huh7 cells (A) treated with sovaldi-chitosan nanoparticles (100 μM) versus (B) untreated cells. There is no obvious harmful effect due to cells treatment with SCNPs**

### 3.3. Effect of SCNPs on the Integrity of genomic DNA

DNA fragmentation analysis was carried out with gel electrophoresis to investigate the effect of sovaldi and SCNPs on the genomic integrity of DNA. **Figure 6** shows that the cellular DNA contents of the treated cells was not different from the control after 24 hr exposure time.

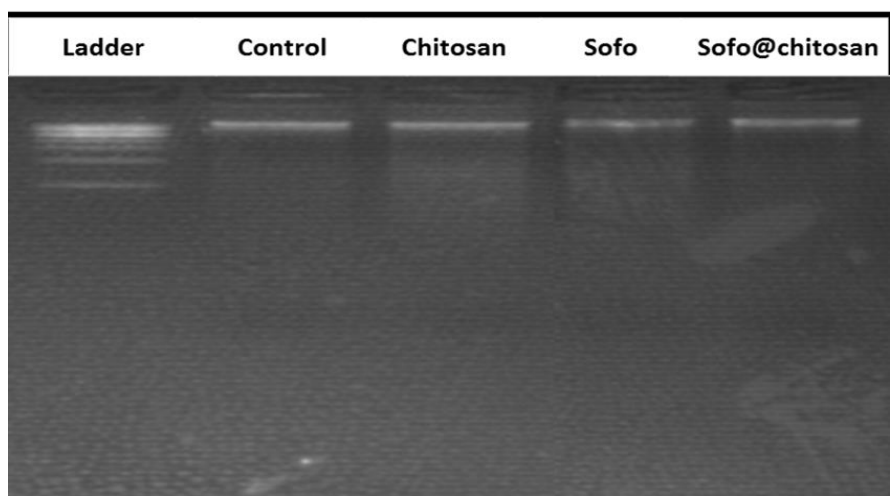

**Figure 6: EB-stained gel electrophoresis of DNA extraction from Huh7 cells, control (untreated cells), CNPs treated cells (chitosan), Sovaldi treated cells (Sofosbuvir), and SCNPs treated cells (Sofo @ chitosan). No differences of cell DNA content was identified among treated and untreated cells.**

### 3.4.    Viral replication in Huh7 cells and antiviral activity of SCNPs

Huh7 cells were incubated with HCV-4 infected human serums for 24 h, 48 h, and 72 h. The total cellular RNA was then extracted from both treated and untreated cells and was subjected to quantitative real-time PCR assay. Our results showed that, the intracellular viral load peaked after 24 h post infection as presented in **Table 2**. Therefore, we used this incubation time for our preliminary investigation for the antiviral activity of the SCNPs where Huh7 cells were incubated with SCNPs (100 µM, 53 mg/ml) for 24h. This resulted in a complete disappearance of HCV RNA compared to the untreated infected cells as presented in **Table 3**.

**Table 2: Viral inoculation and detection of HCV RNA at different time intervals by quantitative real time PCR assay**

| Time interval (h) | Viral Load (IU/mL) |
|---|---|
| Uninfected Huh7 cells (24) | Below detection limit |
| Post infection (24) | $2 \times 10^6$ |
| post infection (48) | $5 \times 10^5$ |
| post infection (72) | $9.6 \times 10^3$ |

**Table 3.   Antiviral monitoring of Huh 7 cells treated with Sovaldi - Chitosan nanoparticles for 24 h using quantitative PCR assay**

| Sample | CT value | Viral load (IU/ml) |
|---|---|---|
| HCV serum | 28.42 | $24 \times 10^4$ |
| Control | 33.38 | $7 \times 10^3$ |
| Sovaldi | below detection limit | <50 |
| SCNPs | below detection limit | <50 |

CT value means cycle threshold values of RT-PCR.

Control, cells infected with HCV-4 but were not treated with Sovaldi or SCNPs

### 3.2.2.   In silico docking study of HCV-4 NS3, NS5A and NS5B Polymerase with Sovaldi and SCNPs

The in-silico docking study was performed to identify the interactions of Sovaldi and SCNPs with HCV-4 NS3, NS5B and NS5A based on MolDock score followed by Hydrogen Bond Interaction (H-bond) score. The binding affinity of Sovaldi, at the active site, showed the best docking score for NS5A polymerase, MolDock score (-167) where Sovaldi formed 8 H-bonds, one H-bond with each of Glu 116 and Arg 157 and two H-bonds with each of Ala 162, Glu 116 and Lys 164. The interaction of NS5A polymerase and SCNPs involved formation of 5 H-Bonds (One H-Bond with each of Arg 160, Leu 149 and Gly96 and (two h-bonds with Trp111) and the MolDock score was -131.

However, SCNPs displayed a higher binding affinity for the active site of NS3 protease than Sovaldi, based on the MolDock score where Sovaldi formed 3 H-Bonds with each of Thr 76, Asn 77 and Ser 181 with -127.581 MolDock score. Contrary to SCNPs where the predicted mode of interaction of SCNPs and NS3 protease involved formation of 11 H-bonds, one H-bond with each of Asn 187, His 203, Gln 73, Pro 332, Ser 334, Thr 185 and Val 183. In addition, it forms two H-bonds with both Asn 72 and Thr 331 with MolDock score equivalent to -156.512.

In case of NS5B polymerase, the MolDock score for SCNPs was -154.603 compared to -131.535 for Sovaldi and this is indicative of a better interaction between NS5B and SCNPs than Sovaldi. SCNPs interaction with NS5B (Figure 7) involved formation of 8 H-Bond, One H-bond with each of Asn 291, Asp 319, Cys 316 and Cys 366 and two hydrogen bonds with Asp 318. While Sovaldi formed 3 H bonds, each with Asn 291, Asp 319, Cys 316.

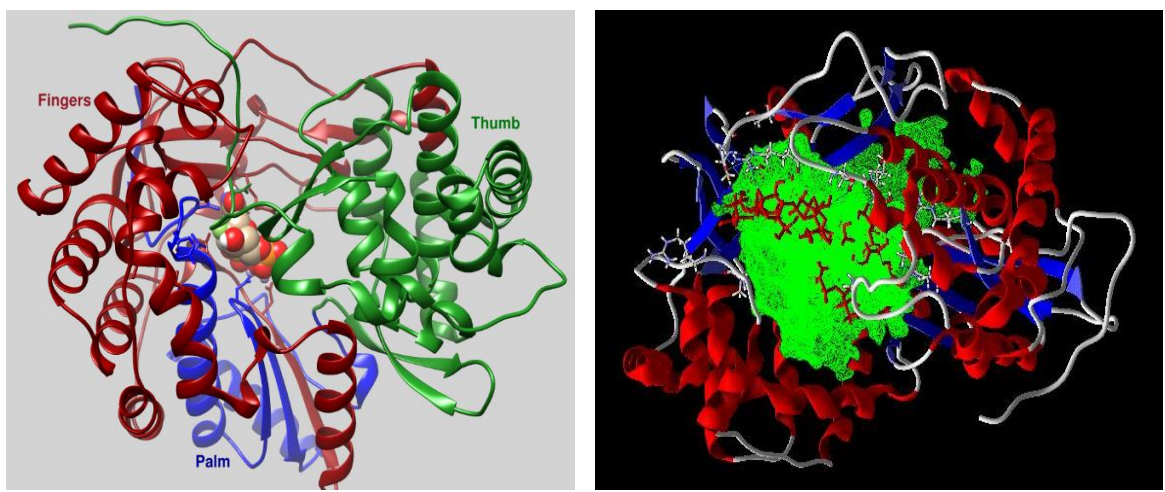

**Figure 7: NS5B polymerase structure and domains (A); Palm, thumb and fingers are colored in blue, green and red respectively. SCNPs interaction with the active site of NS5B polymerase (B).**

## 4.    Discussion

Chronic HCV-4 infection and its complications (cirrhosis, liver cell failure, hepatocellular carcinoma (HCC)) represents one of the major health problem among Egyptian population (21). Currently, Egyptian governmental efforts are focused on fighting such a debilitating infection through research funding, awareness campaigns and scientific meetings. Fortunately, the NS5B polymerase inhibitor (GS-7977; Sovaldi) was approved in Europe (22) and united states (23) for the treatment of chronic hepatitis C. These approvales have enabled the application of Sovaldi in Egypt. Sovaldi may be used in combination with ribavirin either in an interferon-free regimen or in combination with peginterferon-α. The application of Sovaldi depends on the genotype of the HCV and the patient compliance for interferon and/or ribavirin treatments. In contrast, some scientists reported that the viral mutations, therapeutic index, associated side effects, the need for antiviral targeting might be the major obstacles for application of Sovaldi in treatment of HCV (24).

Nanoparticles (NPs) are considered as one of the most effective and novel therapeutic agents due to their potential ability to be tolerated and to target the therapeutic agent to the diseased organ with a concomitant decrease of off-target side effects as described earlier. Therefore, we encapsulated sofosbuvir into chitosan nanoparticles with a future aim of using a lower drug dose for treatment of HCV-4 infected patients due to the ability to target of sofosbuvir into liver.

SCNPs prepared by ionic gelation method for the preparation of SCNPs of a size and zeta potential $137 \pm 34$ nm and $29 \pm 9.6$ mV respectively. The loading percentage of sovaldi into chitosan NPs was calculated as previously described (25) and was 6% with an entrapment efficiency of 80%. The encapsulation of Sovaldi into Chitosan nanoparticles was further confirmed by FTIR analysis where we identified the disappearance of the N-H peak that is indicative of the interaction between chitosan and sofosbuvir through the positively charged ammonium group of chitosan and the lone pair of electrons of sofosbuvir. Further studies are in progress to explore the binding reaction and the kinetic release profile of sovaldi from SCNPs.

We then performed several studies on cellular and molecular levels to prove the safety of the composite before exploring its antiviral activity. The treatment of Huh7 cells with sovaldi or SCNPs show neither toxic effects nor alterations in cellular morphology after incubation of cells with concentrations up to 100 μM (53 mg/ml) for 24 and 48 h. The cellular uptake of the produced SCNPs was similar to previous preparations from our laboratory regarding internalization and distribution in cytoplasm and nucleus (8). Flow cytometric analysis and cellular DNA fragmentation showed no cellular or molecular toxicity. These results are matched with other reported data (26) where human in vitro culture of fibroblasts treated with chitosan nanoparticles showed no cytotoxic effect as revealed by similar cell viability between cells treated with chitosan nanoparticles and control cells (cells cultured in the medium in the absence of chitosan nanoparticles)

Huh7 cells were then infected with HCV-4 positive human sera, at viral loads > 2 x $10^6$ IU/ml. Infected cells were subsequently exposed to either sovaldi or SCNPs. The viral replication (in the absence of sofosbuvir or SCNPs) reached a peak at 24 h. Treating these infected cells by sofosbuvir and SCNPs resulted in complete disappearance of HCV RNA.

Molecular docking study showed that SCNPs form more stable complex with NS3 protease, and NS5B polymerase than Sovaldi as indicated by the docking score values recorded for NS3 and NS5B with SCNPs; -156.512 , -154.603 versus -127.581, -131.535 recorded with Sovaldi. Contrary to NS5A where a more stable complex formed with Sovaldi rather than SCNPs as revealed by their docking score value; -167 for Sovaldi versus -131 for SCNPs. These data reflected that SCNPs could potentiate the antiviral activity of Sovaldi when encapsulated into chitosan nanoparticles.

The obtained data suggested that the encapsulation of Sovaldi into chitosan nanoparticles would not negatively affect its antiviral activity, but it could be associated with a lower side effect. However, further in-vivo investigations are still under processing to assure the safety of sofosbuvir encapsulated chitosan nanoparticles.

**Conclusion**

Virus infections cause diseases of different severity ranged from mild infection e.g. common cold into life threatening diseases e.g. Human Immunodeficiency virus (HIV), Hepatitis B. Virus infections represent 44% of newly emerging infections. Although there are many efficient antiviral agents, they still have drawbacks due to accumulation at off target organs and developing of virus resistance due to virus mutation. Therefore, developing a delivery system that can selectively target drug into affected organs and avoid off target accumulation would be a highly advantageous strategy to improve antiviral therapy. In this work, we encapsulated Sovaldi, a direct acting antiviral agent into chitosan nanoparticles by ionic gelation method, the size of SCNPs was $137 \pm 34$ nm and the nanoparticles were stable as indicated by zeta potential value, $29 \pm 9.6$ mV and TEM images. The percentage of drug encapsulation and loading was 80% and 6% respectively. These NPs displayed no cytotoxic or genotoxic effects on human liver cancer cells, Huh7 cells after exposure to a concentration up to 100 μM (53 mg/ml) for 24 h. Both Sovaldi and SCNPs showed potent anti HCV-4 activities at 100 μM (53 mg/ml). Molecular docking study showed that SCNPs have a higher binding affinity to NS5B and NS3 and a lower binding affinity to NS5A than sovaldi. This is indicative of the better antiviral activity that might be offered by SCNPs compared to Sovaldi. Although our results are promising for the application of SCNPs as a novel delivery system for DAA, several studies are still required to explore the antiviral activity on the replicating system, the minimum inhibitory concentrations and the optimum incubation time of the virus with human cells, *in vitro, to* fine-tune the produced SCNPs.

## 5. Acknowledgment

We would like to thank the Science and Technology Development Fund (STDF), Cairo, Egypt for funding this work, Grant No.15108 and Gilead company for providing Sovaldi as a gift.

## 6. Ethical approval

The research has been approved by the National Cancer Institute-Cairo University, IRB

IRB00004025 Organization No. IORG0003381

## 7.Conflict of interest statement

No conflict of interest

## 8. References

1.    Gomaa A, Allam N, Elsharkawy A, El Kassas M, Waked I. Hepatitis C infection in Egypt: prevalence, impact and management strategies. Hepatic Med Evid Res [Internet]. 2017;9:35–6. Available from: https://www.dovepress.com/corrigendum-hepatitis-c-infection-in-egypt-prevalence-impact-and--peer-reviewed-article-HMER

2.    Irwin KK, Renzette N, Kowalik TF, Jensen JD. Antiviral drug resistance as an adaptive process. Virus Evol [Internet]. 2016;2(1):1–10. Available from: https://academic.oup.com/ve/article-lookup/doi/10.1093/ve/vew014

3.    Chawla A, Wang C, Patton C, Murray M, Punekar Y, Ruiter A de CS. A Review of Long-Term Toxicity of Antiretroviral Treatment Regimens and Implications for an Aging

Population. Infect Dis Ther [Internet]. Springer Healthcare; 2018;7(2):183–95. Available from: http://gateway.proquest.com/openurl?ctx_ver=Z39.88-2004&res_id=xri:pqm&req_dat=xri:pqil:pq_clntid=48207&rft_val_fmt=ori/fmt:kev:mtx:journal&genre=article&issn=2193-6382&volume=7&issue=2&spage=183%0Ahttp://europepmc.org/search?query=(DOI:10.1007/s40121-018

4.   Manns MP, McHutchison JG, Gordon SC, Rustgi VK, Shiffman M, Reindollar R, Goodman ZD, Koury K, Ling MH, Albrecht JK GI. Peginterferon alfa-2b plus ribavirin compared with interferon alfa-2b plus ribavirin for initial treatment of chronic hepatitis C: a randomised trial. Lancet. 2001;358(9286):958–65.

5.   Fried MW, Shiffman ML, Reddy KR, Smith C, Marinos G, Gonçales Jr FL, Häussinger D, Diago M, Carosi G, Dhumeaux D CA. Peginterferon alfa-2a plus ribavirin for chronic hepatitis C virus infection. N Engl J Med. 2002;347(13):975–82.

6.   Varshosaz J, Farzan M. Nanoparticles for targeted delivery of therapeutics and small interfering RNAs in hepatocellular carcinoma. World J Gastroenterol. 2015;21(42):12022–41.

7.   Cover NF, Lai-Yuen S, Parsons AK KA. Synergetic effects of doxycycline-loaded chitosan nanoparticles for improving drug delivery and efficacy. Int J Nanomedicine. 2012;7:2411–9.

8.   Loutfy SA, El-Din HM, Elberry MH, Allam NG, Hasanin MT AA. Synthesis, characterization and cytotoxic evaluation of chitosan nanoparticles: in vitro liver cancer model. Adv Nat Sci Nanosci Nanotechnol. 2016;7(3):035008.

9.   Zheng Y, Su C, Zhao L, Shi Y. Chitosan nanoparticle - mediated co - delivery of shAtg - 5

and gefitinib synergistically promoted the efficacy of chemotherapeutics through the modulation of autophagy. J Nanobiotechnology. BioMed Central; 2017;1–11.

10. Ibrahim AH. Antiviral Activity of Chitosan Nanoparticles Encapsulating Curcumin Against Hepatitis C Virus Genotype 4a in Human Hepatoma Cell Lines. Int J Nanomedicine. 2020;(April).

11. Loutfy SA, Mohamed MB, Abdel-Ghani NT, Al-Ansary N, Abdulla WA, El-Borady OM, Hussein Y EM. Metallic nanomaterials as drug carriers to decrease side effects of chemotherapy (in vitro: Cytotoxicity study). J Nanopharmaceutics Drug Deliv. 2013;1(2):138–49.

12. Zhu ZJ, Ghosh PS, Miranda OR, Vachet RW RV. Multiplexed screening of cellular uptake of gold nanoparticles using laser desorption/ionization mass spectrometr. J Am Chem Soc. 2008;130(43):14139–43.

13. T M. Rapid colorimetric assay for cellular growth and survival: application to proliferation and cytotoxicity assays. J Immunol Methods. 1983;16(65(1-2)):55–63.

14. R N. DNA measurement and cell cycle analysis by flow cytometry. Curr Issues Mol Biol. 2001;3:67–70.

15. Gopinath P, Gogoi SK, Sanpui P, Paul A, Chattopadhyay A GS. . Signaling gene cascade in silver nanoparticle induced apoptosis. Colloids Surfaces B Biointerfaces. 2010;77(2):240–5.

16. Ohno O, Mizokami M, Wu RR, Saleh MG, Ohba KI, Orito E, Mukaide M, Williams R LJ. New hepatitis C virus (HCV) genotyping system that allows for identification of HCV genotypes 1a, 1b, 2a, 2b, 3a, 3b, 4, 5a, and 6a. J Clin Microbiol. 1997;35(1):201–7.

17. Hu Y, Shahidi A, Park S, Guilfoyle D HI. Detection of extrahepatic hepatitis C virus

replication by a novel, highly sensitive, single-tube nested polymerase chain reaction. Am J Clin Pathol. 2003;119(1):95–100.

18.   Abdelwahab S, Rewisha E, Hashem M, Sobhy M, Galal I, Allam WR, Mikhail N, Galal G, El-Tabbakh M, El-Kamary SS WI. Risk factors for hepatitis C virus infection among Egyptian healthcare workers in a national liver diseases referral centre. Trans R Soc Trop Med Hyg. 2012;106(2):98–103.

19.   El-Awady MK, Tabll AA, El-Abd YS, Bahgat MM, Shoeb HA, Youssef SS  et al. HepG2 cells support viral replication and gene expression of hepatitis C virus genotype 4 in vitro. World J Gastroenterol. 2006;12(30):4836–42.

20.   Zekri AR, Bahnassy AA, Hafez MM, Hassan ZK, Kamel M, Loutfy SA, Sherif GM, El-Zayadi AR DS. Characterization of chronic HCV infection-induced apoptosis. Comp Hepatol. 2011;10(1):1.

21.   Miller FD A-RL. Evidence of intense ongoing endemic transmission of hepatitis C virus in Egypt. Proc Natl Acad Sci. 2010;107(33):14757–62.

22.   Agency EM. Sovaldi (sofosbovir): EU summary of product characteristics. 2014.

23.   Inc GS. Sovaldi[TM] (sofosbuvir) tablets, for oral use: US prescribing information. 2013.

24.   Keating GM VA. Sofosbuvir: first global approval. Drugs. 2014;74(2):273–82.

25.   Amjadi I, Rabiee M HM. Anticancer activity of nanoparticles based on PLGA and its co-polymer: in-vitro evaluation. Iran J Pharm Res. 2013;12(4):623–34.

26.   Liu H GC. Preparation and properties of ionically cross-linked chitosan nanoparticles. Polym Adv Technol. 2009;20(7):613–9.